\documentclass[graybox]{svmult}

\usepackage{mathptmx}       
\usepackage{helvet}         
\usepackage{courier}        
\usepackage{type1cm}        
\usepackage{makeidx}         
\usepackage{graphicx}        
\usepackage{multicol}        
\usepackage[bottom]{footmisc}

\makeindex             

\begin{document}

\title*{INSA scientific activities in the space astronomy area}
\author{Ricardo P\'erez Mart\'inez and Miguel S\'anchez Portal}
\institute{Ricardo P\'erez Mart\'inez \at ESAC/INSA  \email{ricardo.perez@sciops.esa.int},
 Miguel S\'anchez Portal \at ESAC/INSA  \email{miguel.sanchez@sciops.esa.int}}
%
%
\maketitle

\abstract*{Support to Astronomy operations is an important and long-lived activity within INSA. Probably the best known (and traditional) INSA activities are those related with real-time spacecraft operations: Ground station maintenance and operation (Ground station engineers and operators); spacecraft and payload real-time operation (spacecraft and instruments controllers); computing infrastructure maintenance (operators, analysts) and	general site services.In this paper, we'll show a different perspective, probably not so well-known, presenting some INSA recent activities at the European Space Astronomy Centre (ESAC) and NASA Madrid Deep Space Communication Complex (MDSCC) directly related to scientific operations. Basic lines of activity involved include: Operations support for science operations; system and software support for real time systems;	technical administration and IT support; R \& D activities, radioastronomy (at MDSCC and ESAC) and scientific research projects. This paper is structured as follows: first, INSA activities in two ESA cornerstone astrophysics missions, XMM-Newton and Herschel, will be outlined. Then, our activities related to Science infrastructure services, represented by the Virtual Observatory (VO) framework and the Science Archives development facilities are briefly shown. Radio Astronomy activities will be described afterwards, and finally, a few research topics in which INSA scientists are involved will be also described.
}

\abstract{Support to Astronomy operations is an important and long-lived activity within INSA. Probably the best known (and traditional) INSA activities are those related with real-time spacecraft operations: Ground station maintenance and operation (Ground station engineers and operators); spacecraft and payload real-time operation (spacecraft and instruments controllers); computing infrastructure maintenance (operators, analysts) and	general site services.In this paper, we'll show a different perspective, probably not so well-known, presenting some INSA recent activities at the European Space Astronomy Centre (ESAC) and NASA Madrid Deep Space Communication Complex (MDSCC) directly related to scientific operations. Basic lines of activity involved include: Operations support for science operations; system and software support for real time systems;	technical administration and IT support; R \& D activities, radioastronomy (at MDSCC and ESAC) and scientific research projects. This paper is structured as follows: first, INSA activities in two ESA cornerstone astrophysics missions, XMM-Newton and Herschel, will be outlined. Then, our activities related to Science infrastructure services, represented by the Virtual Observatory (VO) framework and the Science Archives development facilities are briefly shown. Radio Astronomy activities will be described afterwards, and finally, a few research topics in which INSA scientists are involved will be also described.
}

\section{ESA Astrophysics Missions (ESAC)}
\label{sec:1}
\subsection{XMM Newton}

Many celestial objects generate X-rays in extremely violent processes (supernovae, quasars and active galactic nuclei, gamma-ray bursts, etc.). But Earth's atmosphere blocks out these X-rays, and only by placing X-ray detectors in space can such sources be detected, pinpointed and studied in detail. 

XMM-Newton is ESA's second \textit{cornerstone} mission. It carries three very advanced X-ray telescopes, each containing 58 high-precision concentric mirrors, delicately nested to offer the largest collecting area possible to catch the elusive X-rays. These Mirror Modules allow XMM-Newton to detect millions of sources, far more than any previous X-ray mission. The scientific payload includes six instruments: the three European Photon Imaging Cameras (EPIC), MOS-1, MOS-2 and PN are sensitive devices for imaging and non-dispersive spectroscopy, while the two Reflection Grating Spectrometers (RGS), RGS-1 and RGS-2 are high resolution X-ray grating spectrographs. The X-ray payload is complemented by the Optical Monitor (OM), a 30\,cm telescope for imaging and low-resolution spectroscopy working in the visible and ultraviolet range.

INSA has been participating in the XMM-Newton Science Operations Centre (SOC) since 1999; besides the well-known role of the Instrument Controllers, a number of activities have been carried out, including real-time and off-line activities. INSA personnel are currently in charge of the maintenance of the six instruments on board (monitoring, on board software development, design and maintenance of flight operation and recovery procedures, assistance to critical operations and general engineering support), as well as specific involvement in the calibration teams and in the  scientific support group, dealing with the necessities and requests from the science community and the instrument teams. We are also involved in the development of the data reduction software and the optimization and development of scientific pipelines in Grid architectures. The activities related to the XMM/Newton simulator (thermal aspects, power management, operation tests) are also carried out by our team. Other off-line activities include the coaching of research fellows and trainees and the maintenance of the XMM/Newton public gallery of scientific highlights of the mission. 

\subsection{Herschel Space Observatory}
\label{sec:2}
Herschel, short for the \textit{Herschel Space Observatory} is the fourth cornerstone mission in the European Space Agency (ESA) science programme. It will perform imaging photometry and spectroscopy in the far infrared and submillimetre part of the spectrum, covering approximately the 55-672 µm range. It is the only space facility dedicated to that portion of the electromagnetic spectrum. Its vantage point in space (the Second Lagrange Point of the Sun-Earth system, L2) provides several decisive advantages, including a low and stable background and full access to this part of the spectrum.

Herschel has the potential of discovering the earliest epoch proto-galaxies, revealing the cosmologically evolving AGN-starburst symbiosis, and unraveling the mechanisms involved in the formation of stars and planetary system bodies. The key science objectives emphasise specifically the formation of stars and astrochemistry, and solar system studies.

Herschel will carry a 3.5m diameter passively cooled telescope (the largest space telescope to date!). The science payload complement - two cameras/medium resolution spectrometers (PACS and SPIRE) and a very high resolution heterodyne spectrometer (HIFI) - will be housed in a superfluid helium cryostat.

Herschel will be launched during 2009. Once operational Herschel will offer a minimum of 3 years of routine observations; roughly two thirds of the available observing time is open to the general astronomical community through a standard competitive proposal procedure.

INSA is working in the Herschel Science Centre at ESAC since 2005, providing a number of services grouped in three major items: Science support within the project scientist team (definition of the pointing modes, calibration strategies, pointing and auxiliary products); payload operations (maintenance of the instrument aligment data, mission planning, assitance to the call for porposal porcess, including support to the community and to the Time Allocation Comite); and off-line data management and exploitation (Systematic Product Generation application, Quality Control Browser, Support to the organisation of events (workshops, meetings) and coaching of students and research fellows).

\section{ESA Science Infrastructure Services (ESAC)}
\label{subsec:2}
\subsection{Virtual Observatory}
The Virtual Observatory is an international astronomical community-based initiative. It aims to allow global electronic access to the available astronomical data archives of space and ground-based observatories, sky survey databases. It also aims to enable data analysis techniques through a coordinating entity that will provide common standards, wide-network bandwidth, and state-of-the-art analysis tools.
The International Virtual Observatory Alliance (IVOA) was formed in June 2002 with a mission to \textit{facilitate the international coordination and collaboration necessary for the development and deployment of the tools, systems and organizational structures necessary to enable the international utilization of astronomical archives as an integrated and interoperating virtual observatory.} The IVOA now comprises 16 VO projects from Armenia, Australia, Canada, China, Europe, France, Germany, Hungary, India, Italy, Japan, Korea, Russia, Spain, the United Kingdom, and United States.
ESAC is the center for Spaced Based Virtual Observatory. INSA staff participates in the ESA VO project in a number of activities that include collaborating in the development of the tool VOSpec for handling spectra and participating in the VO interoperability working groups where VO standards are discussed and defined.

\subsection{Development of Science Archives }

The vast amounts of scientific data obtained during a Space Science mission have a much longer lifetime than the satellite mission itself. The data are archived and made freely accessible on-line to the world scientific community, and these archives are frequently a mine of unexpected discoveries. They allow researchers to study, for instance, the evolution of a certain celestial object with time, or its appearance at different wavelengths as observed by different telescopes. The archives of ESA's Astronomy (ISO, XMM-Newton, Integral and Herschel in a near future) and Solar System missions (Mars Express, SMART-1, Rosetta, Huygens, Venus Express and Giotto) are available in ESAC's state-of-the-art archival system.
Within the development of the ESA/ESAC science archives, INSA collaborates in several activities, including: the design, development, deployment and maintenance of astronomical data repositories (databases/data products repository) as well as Middle-Tier Java applications to distribute Science Archive data and the management of the Configuration Control System.

\section {Radio Astronomy (MDSCC and ESAC)}
Radio Astronomy observations started in MDSCC when the Deep Space Exploration began and the DSN Network was pioneer in performing transcontinental Very Long Baseline Interferometry -VLBI- observations. The NASA Deep Space Network (DSN) antennae are excellent radio telescopes because of their large apertures and the performance of their refrigerated low noise amplifiers. Thanks to these characteristics the MDSCC DSS-63 antenna, with 70m diameter, is one of the best single-dish radio telescopes in the world, in terms of its sensitivity and spatial resolution. 
MDSCC participates routinely in the European VLBI Network -EVN- observing sessions and in Global Observations (EVN + Very Long Baseline Array -VLBA-).
Single dish observations (spectroscopy or continuum) are performed as part of the Host Country Program, managed by LAEFF/INTA, that allows the Spanish Astronomical Community to observe with MDSCC radio telescopes (up to 3\% of total time at each antenna). In addition, main VLBI observational areas in which MDSCC has or is currently contributing range from the study of Galactic and extragalactic radio sources (e.g. study of the Galactic Center SgrA* (Alberdi et al. 1999) and  the study of the expansion of the Supernova SN1993J in M81 reported in Marcaide et al. 2002) as well as geodetic studies.

The MDSCC Radio Astronomy Department supports Astronomy Operations at the complex. Among its responsibilities are the maintenance and calibration of the VLBI and spectroscopic equipment and antenna; preparation of local observing procedures and training courses to antenna operators as well as the support to Host Country activities. Development of observational control software
and hardware support are also included.

MDSCC Radio Astronomy Department staff participates in the ASTRID Program and in the Advanced Tracking Observational Techniques Office of the Jet Propulsion Laboratory (JPL/NASA).

\section{R\&D Activities (ESAC staff)}
 Besides the already mentioned activities, INSA personnel plays major roles in long term R\&D projects developed at ESAC. Some of them are outlined below.

\begin{itemize}

	\item	Collaboration with universities: Carlos III, Madrid:
	Development of high-performance file systems for clusters and Grid computing
	Alternative systems for massive data distribution
	\item Collaboration with CETA-CIEMAT: Sun@home: volunteer computation for Sun imaging processing.
	\item	Collaboration with the MAGIC project for the development of scientific software: The project covers the following areas:
	High-energy Astrophysics: detection techniques for hard gamma ray ($E  \geq100$\,GeV).
	Monte-Carlo techniques for simulating atmospheric particle showers initiated by cosmic rays and their detection.
	Grid architecture for massive data processing
	\item Torres Quevedo Fellowship: 
INSA has contracted a scientist within the framework of the Torres Quevedo (TQ) Program.
\end{itemize}
 
\section{Some Research Topics…}
To end this short review paper, a few research topics currently being carried out by INSA staff are outlined. This list is by no means exhaustive but only intended as an example of the activities being developed by our scientists. There are a variety of topics covering a huge range of scales (comets, trans-neptunian objects, brown dwarfs, stars, planetary nebulae, interacting galaxies, active galactic nuclei, clusters of galaxies, gamma-ray bursts, observational cosmology…), as shown below:
\begin{itemize}
	\item		High-mass X-ray binaries. (A. Ibarra).
	\item   Electronic and structural properties of nanotechnology systems: Nanocatalysis, nanocontacts and Scanning Tunneling Microscopy. (J.M. Blanco)
	\item   Statistics of Relativistic Broadened Fe K-alpha Lines in Active Galactic Nuclei (AGN). (I. de la Calle)
	\item		Development of an automatic method for the calculation of chemical abundances (a semi-automatic pipeline). (B. Gonz\'alez-Garc\'ia)
	\item		Minor planets, comets, satellites of giant planets. (M. Kidger) 
	\item		Variability of blazars. (M. Kidger)
	\item		AGNs and interacting galaxies. (N. Loiseau)
	\item		GRB Expansion Model: Dynamics and Emission models of afterglows: SEDs and Light Curves. Magnetic Field and other physical parameters deviation (e.g. matter and electron density). Polarisation. (A. Llorente)
	\item		From Asymptotic Giant Branch to Planetary Nebula Evolution:	Formation of Circumstellar envelopes (gas and dust). Chemical evolution of gas and dust grains. Solid state features in IR spectra Development of evolutive sequences of AGB to PN transition sources. (J.V. Perea)
	\item		Galaxy clusters:	Study and characterization of galaxy cluster population: morphology, star formation history. Comparision with field galaxies. Usage of the gravitational lensing effect in Clusters to study the background sources. (R. P\'erez)
	\item		Multiwavelength study of the AGN population in deep surveys:	X-ray properties (X/O, hardness ratios, etc.). Properties of optical counterparts (morphology, photo-z, etc.). NIR-FIR follow-up. (M. S\'anchez)
	\item		Optical properties of high-energy detected sources. (M. S\'anchez)
	\item		A TF study of SF tracers in galaxy clusters (M. S\'anchez, \& R. P\'erez)

\end{itemize}

%
%
%

\end{document}